\documentclass[showkeys,showpacs,twocolumn,nofootinbib]{revtex4} 

\usepackage[russian,english]{babel}
\usepackage{amsmath,amssymb}
\usepackage{graphicx}
\usepackage{wrapfig}

\usepackage[T2A]{fontenc}
\usepackage[cp1251]{inputenc}

\usepackage{graphicx}
\usepackage{color}
\begin{document}

\title[THE INFLUENCE OF SCREENING EFFECTS]
{THE INFLUENCE OF SCREENING EFFECTS ON THE GRAIN CHARGE IN A THERMAL DUSTY PLASMA}%
\author{D.Yu.~Mishagli}
\affiliation{Department of Theoretical Physics, Faculty of Physics,\\
Odessa I.I. Mechnikov National University,\\
 2 Dvoryans'ka Str., Odessa 65026, Ukraine}
\email{mishagli@onu.edu.ua}

\begin{abstract}
The influence of an inhomogeneous screened electric field on the
charging of dust grains in a thermal dusty plasma is studied. The
electric field of charged grains is considered within the cell
approach, where the problem is reduced to a one-particle one. Within
the model of quasichemical equilibrium, which is generalized to the
case of an inhomogeneous screened electric field, we obtain the
value of mean charge of dust grains, the distribution function of
grains over charges, and the variance of this distribution. We also
give the criterion of an inhomogeneity $g(z)$ of the electric field
and show that  the influence of screening effects can be neglected
in the case of the rarefied subsystem of dust grains (for $r_{\text
c}/r_{\text p} \gg 1$) and in the case of small-radius grains
($r_{\text p} \sim 10^{-5}$~cm).
\end{abstract}


\maketitle

\section{Introduction}
The thermal dusty plasma characterized by the equality of the temperatures of all
components  occurs mainly under the terrestrial conditions in the combustion products of
various fuels \cite{shukla}. One of the important problems of the theory of a dusty
plasma is the problem of determination of the charge of dust grains \cite{fortov_ufn,morfill_rev}.
This problem is solved with the use of various methods considering
its specificity with that or other degree of adequacy. Below, we discuss the most typical
works.

In work \cite{einb}, the existence of an analogy
between the process of thermoionization of dust grains and
the process of ionization of atoms was used for the determination
of the charge of dust grains. The charge of atoms at the ionization varies by at most
several values of electron charge. Therefore, it is possible to neglect the screening effect
and to  consider that the electrons pass a sufficiently large (infinite) path, by leaving the ions
emitting them. The same approach was applied in
work \cite{einb} to the determination of the charge of identical dust grains
of radius $r_{\text p}$. Within this method, the analog of the ionization energy
$I$ in the Saha equation is the work function of electrons leaving the surface of grains, which
have charge $z$: $I \Rightarrow W_0 + z e^2 / r_{\text p}$, where $W_0$ is the work function
for the surface of neutral grains, and the term $z e^2 / r_{\text p}$
($z$ is the charge in units of the electron charge $e$) has meaning of the work needed
to move an electron from the surface of a grain to infinity. It was shown that,
under the made assumptions, the mean charge $Z$ of dust grains
for $Z \gg 1$ is given by the expression
\begin{equation}\label{eq:einb}
Z = \frac {k T r_{\text p}} {e^2} \ln \frac {n_{\text{es}}} {n_{\text e}} + \frac12,
\end{equation}
where $n_{\text{es}} = 2(2 \pi m_{\text e} k T / h^2)^{3/2} \exp(- W_0/k T)$ is
the equilibrium density of electrons near the surface of an emitting grain, and
$n_{\text e}$ is the mean density of electrons in a plasma.

Within the approach
developed in~\cite{einb}, which will be called quasichemical, works \cite{smith,armus1,armus2}
considered the possibility to form
a negative  charge and constructed the distribution function over charges of grains.
It was shown that the coefficient $k T r_{\text p} / e^2$ is the distribution variance.
We note that a small number of dust grains can be negatively charged, but, on the average,
the dust grains in a plasma of combustion products
have a positive charge.

In work \cite{armus3}, the charge of dust grains of the same size
$r_{\text p}$ is determined from the  condition of balance of
currents: the thermoemission current is balanced by the current of
electrons, which recombine with charged dust grains. The
thermoemission current is determined, like that in \cite{einb}, by
the Richardson--Dushman equation. It is assumed that  all electrons
colliding with a dust  grain recombine. The recombination rate is
estimated in the approximation of binary  collisions as $- \alpha
n_{\text e} n_{\text p}$, where $\alpha = \sigma v$, $\sigma$ is the
cross-section of a dust grain ($\sigma \approx \pi r_{\text p}^2$),
$v$ is the mean thermal velocity of relative motion, and $n_{\text
p}$ is the mean density of dust grains in a plasma. The balance
equation for flows allows one to establish the connection between
the mean charge $Z$ and the mean density of electrons $n_{\text e}$:
$Z = kT r_{\text p} / e^2 \, \ln n_{\text{es}} / n_{\text e}$. In
fact, this result coincides with $\eqref{eq:einb}$. In the same
work, the value of $n_{\text e}$ was taken from experimental data
(in principle, it can be calculated theoretically), which gave the
estimate of the value of charge. For example, for $n_{\text e} =
3.47 \times 10^{10}$ cm$^{-3}$, $n_{\text p} = 2 \times 10^7$
cm$^{-3},$ $T=3285$~K, and for the grains with the radius $r_{\text
p} = 10^{-4}$~cm, the charge $Z$ was calculated to be 1700.

There is no basic difference between the above-mentioned approaches, which was indicated
in \cite{sodha1,sodha2}, where
the formula of equilibrium constants was obtained by the kinetic way. This was further used in \cite{einb,smith,armus1,armus2}.

Works \cite{samuy1,samuy2,samuy3,samuy4,zimin_etal} present the attempts to consider
the inhomogeneous distribution of electrons near a dust grain and the influence of the
appropriate electric field on the recombination. In \cite{samuy1},
the screening effects were considered in the linear approximation. Work \cite{zimin_etal}
dealt with a separate dust grain, and the distribution of the electric field
in its neighborhood was found as a solution of the Poisson equation in the Debye--H\"{u}ckel approximation.

A certain contradiction is inherent in the above approaches: the introduction of the notion
``the mean density'' is not proper, because it was assumed that the thermoemission electrons
go to infinity.

Therefore, some different approaches were naturally developed. There, it was considered that the thermoemission electrons
remain in a bounded region including the dust
grain \cite{gibson,zhuh_etal,maren1,maren2,dragan_cell}. This region
(we will call it the cell) is electrically neutral.
The electrons can pass from one cell to another one, but the mean number
of electrons in each cell under the given conditions remains fixed.
Such an approach allows one to consider only one single dust grain in a cell,
i.e., the problem becomes one-particle. We note that, in the frame of cell
approaches, the introduction of the mean density of emitted electrons is quite
proper.

The system composed from identical
spherical dust grains, electrons emitted by them, and a buffer
gas, was considered in \cite{gibson}. Each dust grain is located inside an electrically neutral
spherically symmetric cell of radius $r_{\text c}$. The size of a cell is determined from the
geometric reasoning as half the mean distance between dust grains:
\begin{equation}\label{eq:cell}
r_{\text c} = \frac12 \left( \frac{3}{4 \pi n_{\text p}} \right)^{1/3}.
\end{equation}
The dust grains have the same charge $Z e$. The electric field inside a cell
is determined by the numerical solution of the Poisson equation under relevant
boundary conditions. In \cite{gibson}, it was accepted that, on the boundary of a cell,
the potential and the strength of the electric field are zero.
For the cases of weak and strong screenings of a grain by the electron cloud, the
approximation formulas for the electric field potential were obtained. In work
\cite{zhuh_etal}, it was shown that the solution of the electrostatic problem in the case of
the weak screening ($r_{\text D} \gg r_{\text p}$) transfers in \eqref{eq:einb}, if the subsystem
of dust grains is rarified.

In~\cite{maren1,maren2}, the electric field inside a cell was determined by the solution
of a linearized Poisson equation supplemented by two boundary
conditions, which establish: (i) the equality of the electric field strength
on the boundary of a cell to  zero and (ii) the connection of the electric field strength on the surface
of a grain with its charge.

We note that the boundary condition (i) is trivial (it follows from the
Ostrogradskii--Gauss theorem) and gives no new results.
Moreover, the influence of the electric field on
the charging of dust grains was not considered in the proper way.

Work \cite{dragan_cell}, which was also based on the cell approach to the
description of properties of a dusty plasma, used the notion ``the plasma potential''
(``bulk plasma potential'') introduced earlier in \cite{dragan1,dragan2}.
It was proposed to reckon the electric field potential $\phi$ from this
``plasma potential''. The authors commented on such a choice of
the reference point for the potential that,
only in this case, one can set $\phi (r)|_{r\rightarrow\infty} \rightarrow 0$.
In the previous works, ``the plasma potential'' was defined so that
the Boltzmann distribution was consistent with the condition of the electric neutrality of a plasma. This proposition
is not proper, since the process of thermoelectronic emission from an isolated
grain is nonstationary, and the use of the Boltzmann distribution is impossible.

The consideration of the processes of charging and the influence
of the screening effects on them has become especially
actual after the discovery of ordered structures in the thermal dusty plasma
\cite{nefedov_ufn}. This problem was analyzed,
in particular, in works \cite{zagor1,zagor2,zagor3}, where the influence
of screening effects on the charging of a separate grain was studied. In work \cite{zagor1},
the behavior of the screened field of a dust grain,
which was described with a linearized Poisson equation,
was studied by numerical methods. It was shown that the essential deviation from a linear theory
of screening should be expected for grains,
whose radius is of the order of the Debye  one.
The dynamics of the charging of a dust grain in the presence of
external sources of ionization with regard for the photoemission from its
surface was numerically studied in \cite{zagor2}. The value of
charge was determined from the condition of equality of flows. The densities
of electrons and ions in the vicinity of a dust grain and
the charge of a grain as a function of the time were calculated.
In work \cite{zagor3}, the influence of boundaries on the screening of a point-like dust grain was studied.

In addition, the Saha equation became again used as a tool to describe
the properties of a dusty plasma in the recent works \cite{sodha_etal1,sodha_etal2}
(see also \cite{dragan1,dragan2}). This is explained by the simplicity and the physical clearness
of this approach. Therefore, the necessity to comprehensively study the possibility of the use
of the quasichemical approach seems obvious.

In the present work, we study  the value and the variance of the mean charge of dust
grains in the thermal plasma. To this end, we use: (i) the cell
approximation and (ii) an approach based on the quasichemical model
of the charging of dust grains. No buffer gas is considered.

\section{Definition of a Cell and the Electrostatic Field Energy}
Here, we will formulate the basic positions of the cell model and will determine
the energy of the electrostatic field created by a charged dust grain. The
limiting case of a weakly charged grain will be analyzed as well.

\subsection{The cell model of a dusty plasma}
Let us consider the identical dust grains of radius $r_{\text p}$
with the mean charge $Ze$, which are in equilibrium
with electrons emitted by them.
We assume that it is possible to separate an electrically neutral cell around each grain.
The cell radius $r_{\text c}$ is given by relation \eqref{eq:cell}.
The condition of electric neutrality of a cell takes the form
\begin{equation}\label{eq:neutral_eq_1}
Z e + 4 \pi \int \limits_{r_{\text p}}^{r_{\text c}} \rho (\mathbf r) r^2 dr = 0,
\end{equation}
where the distance $r$ is reckoned from the center of a grain.

We assume that the distributions of the bulk charge $\rho(\mathbf
r)$ and the potential $\phi(\mathbf r)$ of the electrostatic field
inside a cell have spherical symmetry: $\rho(\mathbf r) \Rightarrow
\rho(r)$ and $\phi(\mathbf r) \Rightarrow \phi(r)$. The distribution
of the potential is described in the approximation of
self-consistent field: the potential $\phi(r)$ satisfies the Poisson
equation, in which the charge density $\rho(r)$ is determined by the
Boltzmann distribution.

We note that the charge density in the vicinity of a charged grain
is not always described by the Boltzmann distribution. For example,
it is not proper if the nonstationary  problems,  involve a change
of the charge of a dust grain in the course of time. In this case,
it is expedient to use the methods of nonequilibrium thermodynamics,
as it was proposed in \cite{zagor4,zagor5}.

In order to solve the problem posed in Introduction, we turn to the
linearized Poisson equation. This is justified by the following
reasoning. The potentials, which  are the solutions of the
linearized and nonlinear Poisson equations, are close for $r_{\text
p} < r < r_{\text p} + r_{\text D}$ and $r_{\text p} + r_{\text D} <
r < r_{\text c}$ ($r_{\text D}$ is the Debye  radius of dust
grains). In the second region, the linearized Poisson equation is
generally quite adequate. The considerable difference between the
potentials is significant only for $r \sim r_{\text D}$. This
circumstance is of importance in the problems, in which the detailed
behavior of the potential is defining. To estimate the mean charge
$Z$ within the method used by us, it is necessary to calculate the
energy of the electrostatic field, which is a functional of the
potential. Therefore, the ``fine'' details of the behavior of the
potential are insignificant.

In the dimensionless form, the linearized Poisson equation reads
\begin{equation}\label{eq:pois}
    \Delta_{\tilde r} \psi (\tilde r) - \frac{1}{\lambda^2} \psi (\tilde r) = 0,
\end{equation}
where $\Delta_{\tilde r}$ is the dimensionless radial part of the Laplace operator.
The dimensionless parameters are as follows:
\begin{equation}\label{eq:dimless_varz}
\tilde r = \frac{r}{r_{\text p}}, \quad \varsigma = \frac{r_{\text c}}{r_{\text p}}, \quad \psi(\tilde r) = 1 + \frac{e \phi(\tilde r)}{kT},
\end{equation}
where $\varsigma$ and $\psi(\tilde r)$ are, respectively, the dimensionless radius
of a cell and the dimensionless potential of the electrostatic field inside a cell,
$e$ is the elementary charge, and $k$ is the Boltzmann constant.

The quantity $\lambda$ is the dimensionless Debye  radius equal to
\begin{equation}\label{eq:deb_len}
\lambda = \sqrt{\frac{1}{\tilde n_{\text e0}}}, \quad \tilde n_{\text e0} = \frac{n_{\text e0}}{n_*}, \quad n_* = \frac{kT}{4 \pi e^2 r_{\text p}^2},
\end{equation}
where $n_{\text e0}$ is the mean density of electrons for $\phi(r) = 0$,
i.e., on the boundary of a cell. We consider that the screening of a grain is realized only
by electrons emitted from its surface.

Equation $\eqref{eq:pois}$ is supplemented by boundary conditions on the boundary
of a cell and on the surface of a dust grain:
\begin{equation}\label{eq:bcond}
\left\{
\begin{aligned}
     &\psi (\varsigma) = 1,
\\
    &\left. \frac{\partial \psi (\tilde r)}{\partial \tilde r} \right|_{\tilde r = 1} = - \frac{Z}{Z_0},
\end{aligned}
\right.
\end{equation}
where $Z_0 = kTr_{\text p}/e^2$ (for grains of a radius of $10^{-4}$~cm
at the temperature $T=3000$~K, $Z_0 \sim 10^2$ by the order of magnitude).
The solution of Eq. $\eqref{eq:pois}$ satisfying the boundary conditions
$\eqref{eq:bcond}$ takes the form
\begin{equation}\label{eq:potential}
    \psi(\tilde r) = \frac{1}{\tilde r}
    \frac{(Z/Z_0) \lambda \sh \frac{\varsigma - \tilde r}{\lambda} + \varsigma \left( \lambda \sh \frac{\tilde r - 1}{\lambda} + \ch
    \frac{\tilde r - 1}{\lambda} \right)}
         {\lambda \sh \frac{\varsigma - 1}{\lambda} + \ch \frac{\varsigma - 1}{\lambda}}.
\end{equation}

It is possible to show that the electric field strength
\[
\tilde E(\tilde r) = \frac{1}{\tilde r} \frac{(Z/Z_0) \ch
\frac{\varsigma - \tilde r}{\lambda} - \varsigma \left( \frac 1
\lambda \sh \frac{\tilde r - 1}{\lambda} + \ch \frac{\tilde r -
1}{\lambda} \right)} {\lambda \sh \frac{\varsigma - 1}{\lambda} +
\ch \frac{\varsigma - 1}{\lambda}} +
\]
\begin{equation}\label{eq:el_field}
+ \frac{1}{\tilde r^2} \frac{(Z/Z_0) \lambda \sh \frac{\varsigma -
\tilde r}{\lambda} + \varsigma \left( \lambda \sh \frac{\tilde r -
1}{\lambda} + \ch \frac{\tilde r - 1}{\lambda} \right)} {\lambda \sh
\frac{\varsigma - 1}{\lambda} + \ch \frac{\varsigma - 1}{\lambda}}
\end{equation}
(here, $\tilde E(\tilde r) = e E(r)r_{\text p} / k T$) on the boundary of an electrically neutral
cell is equal to zero. This is a natural consequence of the
Ostrogradskii--Gauss theorem. Hence, the chosen boundary conditions \eqref{eq:bcond}
are quite suitable for the complete solution of the electrostatic problem, and the
boundary condition $E(r_{\text c}) = 0$ should not be considered instead of $\phi (r_{\text c}) = 0.$
We note that $\tilde E(\varsigma) = 0$ with regard for the dependence of the
Debye radius $\lambda$ on the grain charge $Z$ (see the following subsection).

The energy of the electrostatic field $W_{\text{el}}$ can be calculated in the standard way:
\begin{equation}\label{eq:energy_general}
    W_{\text{el}} = \frac{1}{8 \pi} \int \limits_V \mathbf E^2 dV.
\end{equation}

\begin{figure}
\includegraphics[width=\columnwidth]{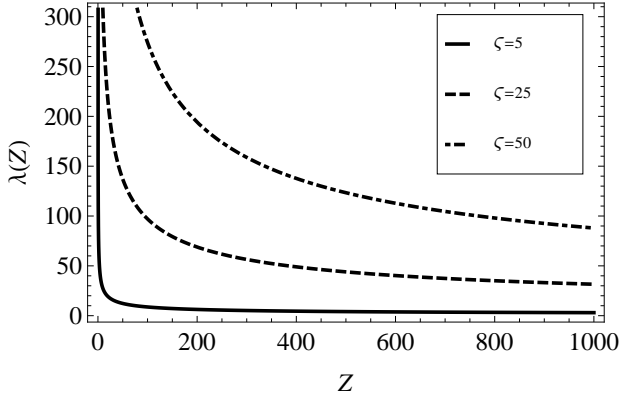}
\vskip-3mm\caption{Debye dimensionless radius $\lambda (Z)$ for
various radii of a cell~$\varsigma$  }\label{fig:deb_len}
\end{figure}

Performing all necessary calculations, we obtain that the dimensionless energy
of the electrostatic field $\tilde W_{\text{el}}(Z) = W_{\text{el}}(Z)/kT$ reads
\begin{equation}\label{eq:energy_dimless_1}
\tilde W_{\text{el}}(Z) = \alpha Z^2/Z_0 - \beta Z + \gamma Z_0,
\end{equation}
where the coefficients $\alpha$, $\beta$, and $\gamma$ are as follows:
\[ \alpha = \frac{\lambda \sh 2 \frac{\varsigma - 1}{\lambda} + 2
\lambda^2 \ch 2 \frac{\varsigma - 1}{\lambda} +
    2 (\varsigma - \lambda^2 - 1)}{8 D},\]
\[\beta = \frac{\varsigma \left[ (\varsigma - \lambda^2 - 1) \sh
\textstyle \frac{\varsigma - 1}{\lambda} +
    \lambda (\varsigma - 1) \ch \frac{\varsigma - 1}{\lambda} \right]}{2 \lambda
    D},\]
\[\gamma = \frac{\varsigma}{8 \lambda^2 D} \bigg[ \lambda \left[
(\varsigma - 4) \lambda^2 + \varsigma \right]
    \sh \textstyle 2 \frac{\varsigma - 1}{\lambda} +\]
\[+2 \lambda^2 (\varsigma - 1 - \lambda^2) \ch 2 \frac{\varsigma - 1}{\lambda} +
\]
\[+2 \left[ (\varsigma^2 + \lambda^2)(\lambda^2 - 1) + \varsigma \right]
\bigg],\]
\begin{equation}\label{eq:alpha-beta-gamma}
    D = \left( \lambda \sh \frac{\varsigma - 1}{\lambda} + \ch \frac{\varsigma - 1}{\lambda} \right)^2.
\end{equation}
These coefficients depend implicitly on the charge
of a dust grain $Z$. To take this fact into account, we determine the dependence $\lambda(Z).$

\subsection{Dependence of the Debye radius on the charge of a dust
grain}\label{sec:debye_length}
It is easy to verify that the equation of electric neutrality $\eqref{eq:neutral_eq_1}$
can be written in the form
\begin{equation}\label{eq:neutral_eq_2}
\frac{Z}{Z_0} - \frac{1}{\lambda^2} \int \limits_{1}^{\varsigma} \psi (\tilde r) \tilde r^2 d\tilde r = 0.
\end{equation}
By substituting  solution  $\eqref{eq:potential}$ in
$\eqref{eq:neutral_eq_2},$ we obtain
\begin{equation}\label{eq:neutral_eq_3}
\frac{Z}{Z_0} = \frac{1}{\lambda} \left[ (\varsigma - \lambda^2) \sh \textstyle \frac{\varsigma - 1}{\lambda} + \lambda (\varsigma - 1) \ch \frac{\varsigma - 1}{\lambda} \right].
\end{equation}

This equation establishes the interrelation between the dimensionless Debye
radius $\lambda$ and the mean charge $Z$ of a grain. The corresponding dependence
for various sizes of a cell $\varsigma$ is presented in Fig.~\ref{fig:deb_len}.
It is seen that, as the size of a cell increases (i.e., as the mean distance
between dust grains increases), the dimensionless Debye radius $\lambda$ becomes much more than 1.

\begin{figure}
\includegraphics[width=\columnwidth]{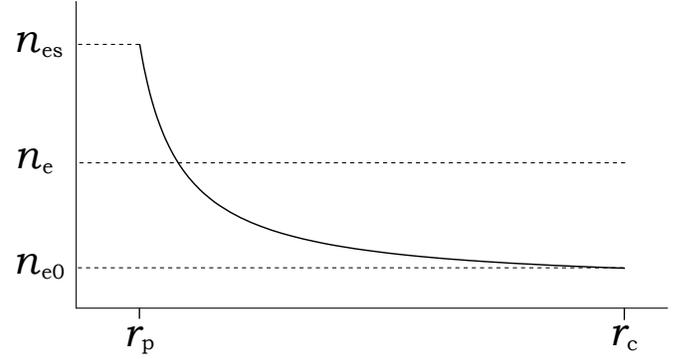}
\vskip-3mm\caption{Schematic distribution of the density of
thermoemission electrons in an electrically neutral cell. Here,
$n_{\text es}$ is the equilibrium density of electrons near the
surface of a grain ($r = r_{\text p}$), $n_{\text e0}$ is the
density of electrons for $\phi=0$ ($r = r_{\text c}$), and $n_{\text
e}$ is the mean density of electrons in a cell
}\label{fig:electron_density}
\end{figure}

Since $\tilde n_{\text e0} = \lambda^{-2}$, Eq. \eqref{eq:neutral_eq_3}
gives also the dependence of the density $n_{\text e0}$ of thermoemission electrons
on the boundary of a cell on the mean charge of a grain $Z$. The knowledge of the value of
$n_{\text e0}$ allows us to determine the mean density $n_{\text e}$
of thermoemission electrons in a cell:
\begin{equation}\label{eq:average_density_1}
    n_{\text e} = \frac{4\pi}{V_{\text c}} \int\limits_{r_{\text p}}^{r_{\text c}} n_{\text e}(r) r^2 dr,
\end{equation}
where $V_{\text c} = 4\pi(r_{\text c}^3 - r_{\text p}^3)/3$ is the cell volume
free of a dust grain. In view of $n_{\text e} (r) \simeq n_{\text e0} \left( 1 + \frac{e\phi(r)}{kT} \right)$, we obtain
\begin{equation}\label{eq:average_density_2}
    n_{\text e}(Z) = \frac{3}{\varsigma^3 - 1} \frac{Z}{Z_0 }n_{\ast} = \frac{3}{4\pi} \frac{Z}{r_{\text c}^3 - r_{\text p}^3}.
\end{equation}
It is clear that the calculated mean density of electrons $n_{\text e}$
in a cell in the limiting case $\varsigma \rightarrow \infty$ should coincide
with $n_{\text e0}$ (see Fig.~\ref{fig:electron_density}), which has meaning, in this case,
of the mean density of electrons in the system.

Equation \eqref{eq:neutral_eq_3} is significantly simplified in the limiting case
of small charges of a dust grain ($Z \rightarrow 0$), which  corresponds to the condition
$\lambda \rightarrow \infty$. In this case, we have a small parameter
$\sqrt{\tilde n_{\text e0}}$ and can expand the right-hand side of
Eq. \eqref{eq:neutral_eq_3} in the series in it. To within $(Z/Z_0)^{3/2},$ we obtain

\begin{equation}\label{eq:small_n}
\tilde n_{\text e0}(Z) \simeq \left( \frac{Z}{a_2 Z_0} \right) \left( 1 - \frac{a_4}{a_2^2} \frac{Z}{Z_0} \right),
\end{equation}
where $a_2 = (\varsigma^3 - 1) / 3$ and $a_4 = (\varsigma^5 - 5 \varsigma^3 + 5 \varsigma^2 - 1) / 30$
are the coefficients of the relevant degrees of $\sqrt{\tilde n_{\text e0}}$. The domain of
applicability of the asymptotic
formula $\eqref{eq:small_n}$ is determined by the inequality $\tilde n_{\text e0} \ll a_2 /a_4$.

Thus, as $\lambda \rightarrow \infty,$ we have the relation
\begin{equation}\label{eq:biglambda}
\lambda(Z) \simeq \left( a_2\frac{Z_0}{Z} \right)^{1/2} \left( 1 + \frac{a_4}{2 a_2^2} \frac{Z}{Z_0} \right),
\end{equation}
which is valid for $\lambda \gg \sqrt{a_2 /a_4}$.

\subsection{Behavior of the potential and the energy of the electrostatic field for large
Debye radii (\boldmath$\lambda \gg 1$)} In this limiting case, the
obtained formulas for the potential \eqref{eq:potential} and the
energy \eqref{eq:energy_dimless_1} of the electrostatic field are
considerably simplified. We now show that, as $\lambda \rightarrow
\infty,$ we arrive at the case considered in Appendix.

Indeed, the condition $\lambda \rightarrow \infty$ is equivalent to $Z \rightarrow 0$.
Therefore, taking \eqref{eq:biglambda} into  account and expanding \eqref{eq:potential}
in the series in $Z$, we obtain \eqref{eq:ap_a_potential}. Such a behavior
of the cell potential was indicated in work~\cite{zhuh_etal}.

If we restrict ourselves by the first term in the expansion, then the formulas for $\alpha$, $\beta,$ and $\gamma$
become
\[
\alpha \simeq \frac{\varsigma - 1}{2 \varsigma} + \ldots, \quad
\beta \simeq \frac{\varsigma^3 - 3\varsigma + 2 }{2 \varsigma
(\varsigma^3 - 1)} \frac{Z}{Z_0} + \ldots, \]
\begin{equation}
 \gamma  \simeq
\frac{\varsigma^6 - 5\varsigma^3 + 9\varsigma - 5}{10\varsigma
(\varsigma^3 - 1)^2} \left(\frac{Z}{Z_0}\right)^2 + \ldots
\end{equation}
Hence, the electrostatic energy \eqref{eq:energy_dimless_1} in the
approximation $\lambda \rightarrow \infty$ ($Z \rightarrow 0$) is
equal to \eqref{eq:ap_a_energy}. In the limiting case $\varsigma
\rightarrow \infty,$ the energy $W_{\text{el}}$ passes into the
energy of a nonscreened grain: $W_{\text{el}} \Rightarrow (Ze)^2 /
2r_{\text p}$.

\section{The Model of Quasichemical Equilibrium}
In this section, we recall  the main results of works
\cite{einb, smith, armus1, armus2} and generalize the approach
developed in them. For this purpose, we use the results obtained in the previous
section. We will construct the distribution function of grains over charges,
like that in \cite{armus2} but with regard for the performed
corrections, and will determined the variance $Z_{\text D}$ of this distribution.

\subsection{Generalization to the case of a screened field}
It is known \cite{landau_lifshitz} that, from the thermodynamic
viewpoint, the ionization equilibrium is a particular case of the
chemical equilibrium corresponding to simultaneously running
``reactions of ionization.'' These  reactions can be written as
follows:
\begin{equation}\label{eq:collisions}
A_0 = A_1 + e^-, \quad A_1 = A_2 + e^-, \quad \ldots,
\end{equation}
where the symbol $A_0$ means a neutral  atom, $A_1$, $A_2$,~\ldots\,
are atoms ionized one, two, {\it etc.} times, and $e^-$ is an
electron.

Under the assumption that such a model of charging holds also for dust
grains, the authors of works \cite{smith, armus1, armus2} obtained the equilibrium constants
\begin{equation}\label{eq:equilib_const}
K_z = n_{\text{es}} \exp \left( - \frac{z \, e^2}{r_{\text p} k T}
\right).
\end{equation}
As was mentioned above (see Introduction), the factor $z e^2 / r_{\text p}$ in relation \eqref{eq:equilib_const}
represents the additional work made by an electron, by moving
from the surface of a grain with charge $z$
to infinity.

The consideration of the influence of the electric field becomes
essential in a number of cases. For example, for grains of the
radius $r_{\text p} = 10^{-5}$~cm with $Z=100,$ the ratio of the
factor $Ze^2/r_{\text p}$ to the work function $W_0 = 2.75$~eV is
equal to 0.52. We obtain the same value also for grains of the
radius $r_{\text p} = 10^{-4}$~cm with $Z=1000$. Thus, the influence
of the electric field becomes essential (for grains of the radius
$r_{\text p} = 10^{-5}$--$10^{-4}$~cm) for $Z \sim 100 $--$1000$.
Such charges are typical  of a dusty plasma.

The Saha equation $n_{\text p}^{(z)}n_{\text e} / n_{\text p}^{(z-1)} = K_z$
allows one to represent the mean density of dust grains with charge $z$
as $n_{\text p}^{(z)} = \prod_{i=0}^{z-1} K_z / n_{\text e}$
and the mean densities of dust grains $n_{\text p}$ and thermoemission electrons
$n_{\text e}$ as
\begin{equation}\label{eq:einb_densities}
n_{\text p} = \sum\limits_{z =-\infty}^{\infty} n_{\text p}^{(z)},
\quad n_{\text e} = \sum\limits_{z =-\infty}^{\infty} z n_{\text
p}^{(z)}.
\end{equation}
In the  above-mentioned  works, the condition of electric neutrality
$n_{\text e} = Z n_{\text p}$ yields  the  formula for the mean
charge $Z$ of dust grains:
\begin{equation}\label{eq:charge_einb+armus}
Z =
\frac {\sum \limits_{z=-\infty}^{\infty} z\exp \left( -\frac{z(z-1)e^2}{2r_{\text p}kT} + z \ln\frac{n_{\text{es}}}{n_{\text e}} \right)}
      {\sum \limits_{z=-\infty}^{\infty} \exp \left( -\frac{z(z-1)e^2}{2r_{\text p}kT} + z\ln\frac{n_{\text{es}}}{n_{\text e}}\right)}.
\end{equation}

\begin{figure}
\includegraphics[width=\columnwidth]{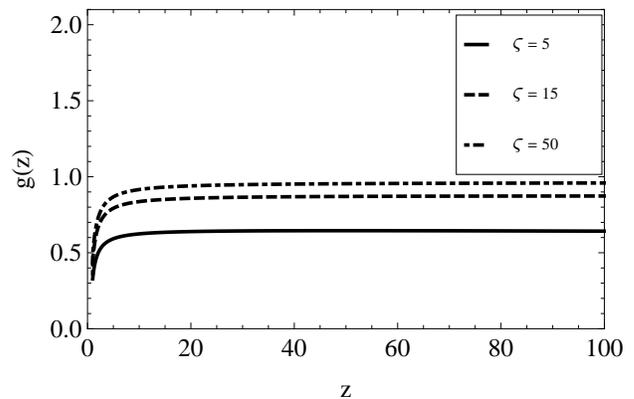}
\vskip-3mm\caption{Function $g(z)$ for various values of the
dimensionless radius of a cell $\varsigma$  }\label{fig:energies}
\end{figure}

Equation \eqref{eq:charge_einb+armus} for $Z$ is not closed, since
the quantity $n_{\text e}$ remains undetermined. In works
\cite{smith, armus1, armus2}, it was taken from experimental data.
In addition, the probability for a grain to obtain a negative charge
is determined, obviously, only by the energy  of the electrostatic
interaction. Therefore, for $z \leqslant 0,$ we should set
$n_{\text{es}} = n_{\text e}$, and we obtain $\ln n_{\text{es}} /
n_{\text e} =0$ in \eqref{eq:charge_einb+armus}.

We now generalize the result in \eqref{eq:charge_einb+armus}, by considering (i)
the existence of a finite domain (electrically neutral cell); (ii) the influence of
screening effects. Then the equilibrium constants $\eqref{eq:equilib_const}$ take the form
\begin{equation}\label{eq:equilib_const_cell}
K_z = n_{\text{es}} \exp \left( - \frac{W_{\text{el}}(z) - W_{\text{el}}(z-1)}{k T} \right),
\end{equation}
where the difference $W_{\text{el}}(z) - W_{\text{el}}(z-1)$ represents the additional work made by
an electron, by moving from the surface of a grain into the cloud of electrons
surrounding a grain.

In Fig.~\ref{fig:energies}, we present the function
\begin{equation}\label{eq:function_g}
\frac{W_{\text{el}}(z) - W_{\text{el}}(z-1)}{z e^2 / r_{\text p}} \equiv g(z)
\end{equation}
for various sizes of a cell $\varsigma$. It is seen from the  figure that, for $\varsigma \gg 1$,
i.e.,  for distances between dust grains to be much more than their radii,
the function $g(z) \rightarrow 1.$ Hence, we may consider that the thermoemission electrons
move from grains to  infinity. Therefore, for $r_{\text c} \gg r_{\text p},$
we may neglect the screening and use the equilibrium constants
\eqref{eq:equilib_const}.

The equilibrium constants \eqref{eq:equilib_const_cell} lead to the result
\begin{equation}\label{eq:charge_einb+cell}
Z =
\frac {\sum \limits_{z=-\infty}^{\infty} z \exp \left[ f(z,Z) \right]}
      {\sum \limits_{z=-\infty}^{\infty} \exp \left[ f(z,Z) \right]},
\end{equation}
where
\begin{equation}
f(z,Z) = \begin{cases}
-\tilde W_{\text{el}}(z),                                             & z \leqslant 0;\\
-\tilde W_{\text{el}}(z) + z \ln\frac{n_{\text{es}}}{n_{\text e}}, & z > 0.
\end{cases}
\end{equation}
Here, the mean density of thermoemission electrons $n_{\text e}$ in a cell
is a function of the mean charge of dust grains $Z$ (see Eq.
\eqref{eq:average_density_2}).

The mean charges of dust grains $\mathrm{CeO_2}$
($W_0 = 2.75$~eV) for various temperatures and sizes of a cell
(or, what is the same, for various mean distances between dust grains)
are given in Tables~\ref{tab:Z_2000}--\ref{tab:Z_2500}.

\begin{table*}[!]
\vspace{3mm} \noindent\caption{ Mean charges of dust grains
calculated by the generalized formula (\ref{eq:charge_einb+cell})
and by formula (\ref{eq:charge_einb+armus}) for various radii of a
cell \boldmath$\varsigma$ and grains \boldmath$r_{\text p}$ at the
temperature \boldmath$T = 2000$~K }\vskip3mm\tabcolsep13.2pt

\noindent{\footnotesize\begin{tabular}{c c c c c c c c c c} \hline
\multicolumn{1}{c|}{$\phantom{\varsigma}$} & \multicolumn{9}{c}{$T = 2000$ K}\\
\cline{2-10}
\multicolumn{1}{c|}{$\varsigma$} & \multicolumn{3}{c}{$r_{\text p} = 10^{-4}$ cm} & \multicolumn{3}{|c}{$r_{\text p} = 5 \times 10^{-5}$ cm} & \multicolumn{3}{|c}{$r_{\text p} = 10^{-5}$ cm}\\
\cline{2-10}
\multicolumn{1}{c|}{$\phantom{\varsigma}$} & \multicolumn{1}{c}{$Z_{\text{cell}}$} & \multicolumn{1}{|c}{$n_{\text e0},$ cm$^{-3}$} & \multicolumn{1}{|c}{$Z_{\infty}$} & \multicolumn{1}{|c}{$Z_{\text{cell}}$} & \multicolumn{1}{|c}{$n_{\text e0},$ cm$^{-3}$} & \multicolumn{1}{|c}{$Z_{\infty}$} & \multicolumn{1}{|c}{$Z_{\text{cell}}$} & \multicolumn{1}{|c}{$n_{\text e0},$ cm$^{-3}$} & \multicolumn{1}{|c}{$Z_{\infty}$}\\
\hline
5   & 757  & $1.46\times10^{12}$ & 427  & 261 & $4.02\times10^{12}$ & 153 & 13  & $2.55\times10^{13}$ & 9   \\
10  & 863  & $2.06\times10^{11}$ & 662  & 338 & $6.46\times10^{11}$ & 263 & 30  & $7.10\times10^{12}$ & 24  \\
15  & 942  & $6.67\times10^{10}$ & 797  & 385 & $2.18\times10^{11}$ & 328 & 40  & $2.84\times10^{12}$ & 35  \\
25  & 1061 & $1.62\times10^{10}$ & 966  & 449 & $5.49\times10^{10}$ & 410 & 54  & $8.24\times10^{11}$ & 50  \\
35  & 1149 & $6.40\times10^{9}$  & 1078 & 495 & $2.20\times10^{10}$ & 465 & 63  & $3.53\times10^{11}$ & 60  \\
50  & 1248 & $2.38\times10^{9}$  & 1196 & 545 & $8.33\times10^{9}$  & 523 & 74  & $1.41\times10^{11}$ & 71  \\
150 & 1582 & $1.12\times10^{8}$  & 1562 & 713 & $4.04\times10^{8}$  & 705 & 107 & $7.57\times10^{9}$  & 106 \\
\hline
\end{tabular}}\label{tab:Z_2000}\vspace*{4mm}
\end{table*}

\begin{table*}[!]
\noindent\caption{ Mean charges of dust grains calculated by the
generalized formula (\ref{eq:charge_einb+cell}) and by formula
(\ref{eq:charge_einb+armus}) for various radii of a cell
\boldmath$\varsigma$ and grains \boldmath$r_{\text p}$ at the
temperature \boldmath$T = 2250$~K }\vskip3mm\tabcolsep13.2pt

\noindent{\footnotesize\begin{tabular}{c c c c c c c c c c} \hline
\multicolumn{1}{c|}{$\phantom{\varsigma}$} & \multicolumn{9}{c}{$T = 2250$ K}\\
\cline{2-10}
\multicolumn{1}{c|}{$\varsigma$} & \multicolumn{3}{c}{$r_{\text p} = 10^{-4}$ cm} & \multicolumn{3}{|c}{$r_{\text p} = 5 \times 10^{-5}$ cm} & \multicolumn{3}{|c}{$r_{\text p} = 10^{-5}$ cm}\\
\cline{2-10}
\multicolumn{1}{c|}{$\phantom{\varsigma}$} & \multicolumn{1}{c}{$Z_{\text{cell}}$} & \multicolumn{1}{|c}{$n_{\text e0},$ cm$^{-3}$} & \multicolumn{1}{|c}{$Z_{\infty}$} & \multicolumn{1}{|c}{$Z_{\text{cell}}$} & \multicolumn{1}{|c}{$n_{\text e0},$ cm$^{-3}$} & \multicolumn{1}{|c}{$Z_{\infty}$} & \multicolumn{1}{|c}{$Z_{\text{cell}}$} & \multicolumn{1}{|c}{$n_{\text e0},$ cm$^{-3}$} & \multicolumn{1}{|c}{$Z_{\infty}$}\\
\hline
5   & 1250 & $2.41\times10^{12} $& 676  & 472 & $7.27\times10^{12}$ & 264 & 36  & $7.01\times10^{13}$ & 23  \\
10  & 1265 & $3.02\times10^{11} $& 955  & 520 & $9.95\times10^{11}$ & 398 & 56  & $1.35\times10^{12}$ & 45  \\
15  & 1326 & $9.38\times10^{10}$ & 1113 & 562 & $3.18\times10^{11}$ & 474 & 68  & $4.81\times10^{12}$ & 59  \\
25  & 1441 & $2.20\times10^{10}$ & 1308 & 627 & $7.66\times10^{10}$ & 570 & 83  & $1.27\times10^{11}$ & 77  \\
35  & 1533 & $8.54\times10^{9}$  & 1436 & 675 & $3.01\times10^{10}$ & 633 & 94  & $5.28\times10^{11}$ & 89  \\
50  & 1641 & $3.14\times10^{9}$  & 1571 & 731 & $1.12\times10^{10}$ & 700 & 105 & $2.01\times10^{11}$ & 102 \\
150 & 2013 & $1.42\times10^{8}$  & 1988 & 918 & $5.20\times10^{8}$  & 907 & 143 & $1.01\times10^{10}$ & 142 \\
\hline
\end{tabular}}\label{tab:Z_2250}\vspace*{4mm}
\end{table*}

\begin{table*}[!]
\noindent\caption{ Mean charges of dust grains calculated by the
generalized formula (\ref{eq:charge_einb+cell}) and by formula
(\ref{eq:charge_einb+armus}) for various radii of a cell
\boldmath$\varsigma$ and grains \boldmath$r_{\text p}$ at the
temperature \boldmath$T = 2500$~K  }\vskip3mm\tabcolsep13.2pt

\noindent{\footnotesize\begin{tabular}{c c c c c c c c c c} \hline
\multicolumn{1}{c|}{$\phantom{\varsigma}$} & \multicolumn{9}{c}{$T = 2500$~K}\\
\cline{2-10}
\multicolumn{1}{c|}{$\varsigma$} & \multicolumn{3}{c}{$r_{\text p} = 10^{-4}$ cm} & \multicolumn{3}{|c}{$r_{\text p} = 5 \times 10^{-5}$ cm} & \multicolumn{3}{|c}{$r_{\text p} = 10^{-5}$ cm}\\
\cline{2-10}
\multicolumn{1}{c|}{$\phantom{\varsigma}$} & \multicolumn{1}{c}{$Z_{\text{cell}}$} & \multicolumn{1}{|c}{$n_{\text e0},$ cm$^{-3}$} & \multicolumn{1}{|c}{$Z_{\infty}$} & \multicolumn{1}{|c}{$Z_{\text{cell}}$} & \multicolumn{1}{|c}{$n_{\text e0},$ cm$^{-3}$} & \multicolumn{1}{|c}{$Z_{\infty}$} & \multicolumn{1}{|c}{$Z_{\text{cell}}$} & \multicolumn{1}{|c}{$n_{\text e0},$ cm$^{-3}$} & \multicolumn{1}{|c}{$Z_{\infty}$}\\
\hline
5   & 1782 & $3.43\times10^{12} $& 933  & 705  & $1.09\times10^{13}$ & 381  & 76  & $9.27\times10^{13}$ & 44  \\
10  & 1681 & $4.02\times10^{11} $& 1255 & 710  & $1.36\times10^{12}$ & 536  & 97  & $1.21\times10^{13}$ & 75  \\
15  & 1719 & $1.22\times10^{11} $& 1434 & 744  & $4.21\times10^{11}$ & 624  & 109 & $3.71\times10^{12}$ & 93  \\
25  & 1827 & $2.79\times10^{10}$ & 1654 & 807  & $9.87\times10^{10}$ & 733  & 125 & $8.48\times10^{11}$ & 115 \\
35  & 1922 & $1.07\times10^{10}$ & 1797 & 858  & $3.82\times10^{10}$ & 804  & 137 & $3.21\times10^{11}$ & 129 \\
50  & 2038 & $3.89\times10^{9}$  & 1949 & 918  & $1.40\times10^{10}$ & 879  & 151 & $1.14\times10^{11}$ & 145 \\
150 & 2447 & $1.73\times10^{8}$  & 2415 & 1125 & $6.36\times10^{8}$  & 1110 & 180 & $1.27\times10^{10}$ & 178 \\
\hline
\end{tabular}}\label{tab:Z_2500}
\end{table*}

First, we calculated the charge $Z$ for given $\varsigma$,
$T,$ and $r_{\text p}$ by formula \eqref{eq:charge_einb+cell}
(column $Z_{\text{cell}}$). In this case, we determined self-consistently
the values of mean density of electrons $n_{\text e}$ in a cell by
formula \eqref{eq:average_density_2}. Then we calculated the values
of charge by the ``old'' formula \eqref{eq:charge_einb+armus}
(column $Z_{\infty}$), where we used the calculated value of $n_{\text e}$.
The computations were performed  with the help of the developed algorithm in language
\texttt{Java} with the use of class \texttt{BigDecimal} from packet \texttt{Math}.

\begin{figure}
\includegraphics[width=\columnwidth]{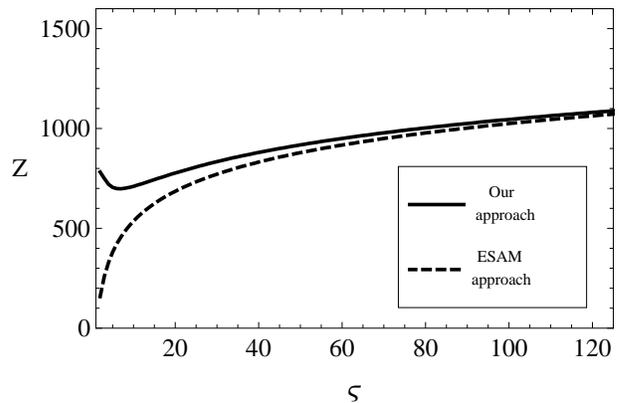}
\vskip-3mm\caption{Mean charge $Z$ of dust grains $\mathrm{CeO_2}$
of the radius $r_{\text p}=5\times10^{-5}$~cm vs the dimensionless
radius of a cell $\varsigma$ at $T=2500$~K
}\label{fig:aver_charge}\vskip3mm
\end{figure}

In Fig.~\ref{fig:aver_charge}, we present the mean charge $Z$ of
dust grains $\mathrm{CeO_2}$ of the radius $r_{\text
p}=5\times10^{-5}$~cm versus the dimensionless radius of a cell
$\varsigma$ at $T=2500$~K. It is seen that the role of the
inhomogeneity is significant for small cells. As the size of a cell
increases, formula \eqref{eq:charge_einb+cell} passes in
\eqref{eq:charge_einb+armus}. In addition, the value of $Z$
decreases for $\varsigma \leqslant 5$ and starts to grow for
$\varsigma > 5.$ We explain such a result by that  the quasichemical
approach considered in the present work is not applicable for very
dense systems ($\varsigma \sim 1$).

\subsection{Distribution of grains over charges}
The undoubtful advantage of the approach under consideration is the
possibility to construct the distribution function of dust grains
over charges like that in work \cite{armus2}, where the equilibrium
constants $\eqref{eq:equilib_const}$ were used. Thus, by assuming
that the charge of grains varies continuously, the following
relation was obtained:
\begin{equation}\label{eq:distrib_fun_1}
\frac{n_p^{(z)}}{n_p} = \frac{1}{\sqrt{2\pi Z_0}} \exp \left( -\frac{z(z-1)}{2Z_0} + z\ln\frac{n_{\text{es}}}{n_{\text e}}\right) \equiv \rho_1 (z).
\end{equation}
The variance of this distribution is $Z_0 = kTr_{\text p}/e^2$.

In the cell approximation, the distribution law for  the charge $z$
of dust grains reads
\begin{equation}\label{eq:distrib_fun_2}
\frac{n_p^{(z)}}{n_p} =
\frac{\exp \left[ f(z,Z) \right]}
     {\sum\limits_{z=-\infty}^{\infty} \exp \left[ f(z,Z) \right]}
                      \equiv \rho_2 (z).
\end{equation}

\begin{figure}
\includegraphics[width=\columnwidth]{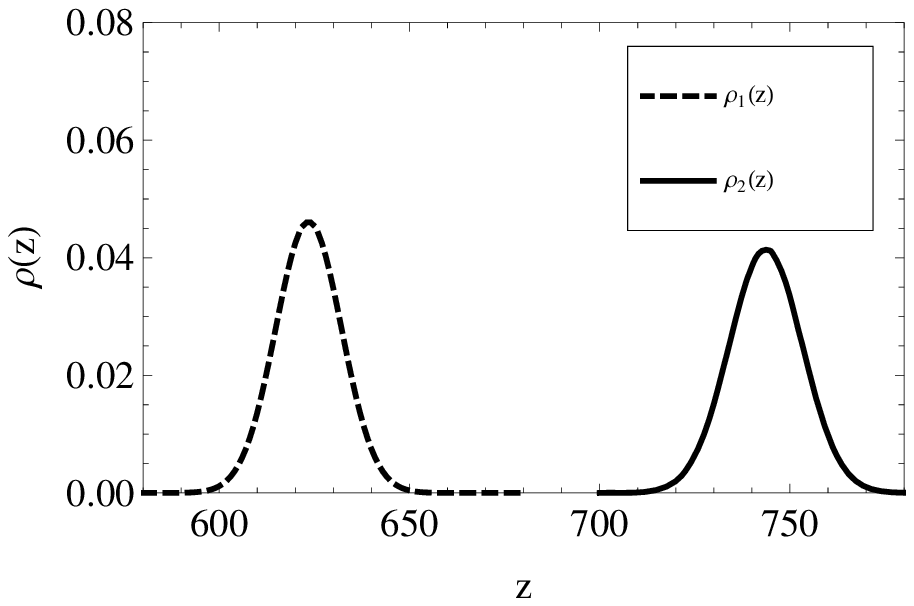}
\vskip-3mm\caption{Distribution functions $\rho_1(z)$ and
$\rho_2(z)$ at $T=2500$~K, $r_{\text p} = 5\times10^{-5}$~cm, and
$\varsigma = 15$  }\label{fig:determ_func}
\end{figure}

In Fig.~\ref{fig:determ_func}, we present the distribution functions
$\rho_1(z)$ and $\rho_2(z)$ at $T=2500$~K, $r_{\text p} =
5\times10^{-5}$~cm, and $\varsigma = 15$. It is seen that the
insignificantly small part of grains is negatively charged.

The variance $Z_{\text D}$ of distribution \eqref{eq:distrib_fun_2} can be determined
in the standard way as
\begin{equation}\label{eq:dispersion}
Z_{\text D} = \mu_2 - \mu_1^2,
\end{equation}
where $\mu_i$ is the $i$-th moment of the distribution, which is
\begin{equation}
\mu_i = \sum\limits_{z = -\infty}^{\infty} z^i \rho_2(z).
\end{equation}
For the same parameters as in Fig.~\ref{fig:aver_charge}, we give the
dependence $Z_{\text D} (\varsigma)$. It is seen that
the variance $Z_{\text D}$ is somewhat different from $Z_0$ for small radii
of a cell, and
$Z_{\text D} (\varsigma) \rightarrow Z_0$ as $\varsigma \rightarrow \infty$.

\begin{figure}
\includegraphics[width=8cm]{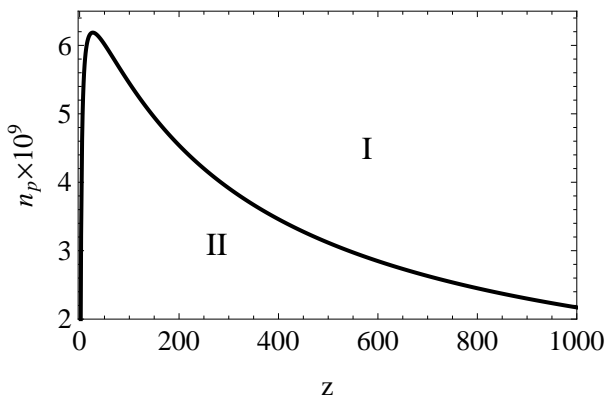}
\vskip-3mm\caption{Line corresponding to the values of $n_{\text p}$
and $z$ (at $T=2500$~K and $r_{\text p} = 5\times10^{-5}$~cm), for
which the condition $g(z) = 0.9$ is satisfied. Regions I and II
correspond to $g(z) < 0.9$ and $g(z) > 0.9$, respectively
}\label{fig:theory_applicability}
\end{figure}

\section{Discussion of Results}

Here, we have presented the results of theoretical studies of the influence
of the screened
electric field on the charging of dust grains in a thermal plasma.
We have applied the following approximations: (i) the cell model of dusty plasma
for the description of the distribution of the electrostatic field; (ii) the quasichemical approach
for the determination of the mean charge of dust grains, distribution of grains
over charges, and variance of this distribution.

The introduction of an electrically neutral cell containing a grain allowed us to: (i)
consider the circumstance that the emitted electrons remain in the vicinity of a dust
grain, rather than move to infinity; (ii) describe satisfactorily
the distribution of the potential of the electrostatic field of a charged grain. Due to the
introduction of a cell, we obtained a closed system of equations
\eqref{eq:neutral_eq_3},
\eqref{eq:average_density_2}, and \eqref{eq:charge_einb+cell} for the determination of the mean charge $Z$.

The influence of the inhomogeneity of the distribution of the electrostatic field is significant for $g(z) < 1$.
In Fig.~\ref{fig:theory_applicability}, we demonstrate the line corresponding to $g(z) = 0.9$
at the temperature $T=3000$~K and the radius of grains $r_{\text p} = 10^{-4}$~cm.
Above this line (region I), we deal with densities $n_{\text p}$ and charges
$z$, for which it is necessary to use the corrected formulas for the determination
of the mean charge of dust grains. Below this line (region II), the influence of
the screening effect can be neglected.

We have shown that the screening causes an increase of the mean charge
of dust grains as compared with the prediction of the theory omitting
the screening effects. As the size of a cell increases, i.e., as the mean
distance between dust grains increases, result
\eqref{eq:charge_einb+cell} is transformed into the well-known
Einbinder--Smith--Arshinov--Musin  formula \eqref{eq:charge_einb+armus}.

For the grains of the radius $r_{\text p} = 10^{-5}$~cm, the differences in the predictions
of the two approaches are slight. This is explained by the fact that the grains of such
sizes have very  small charges ($\sim 1$), and the screening plays no significant role.

In the experiments \cite{fortov_etal} with grains $\mathrm{CeO_2}$
($W_0 = 2.75$~eV), the mean density of grains $n_{\text p}$ varied
in the limits (0.2--5.0)~$ \times 10^7$~cm$^{-3}$, and the
temperature $T$ was changed in the interval (1700--2200)~K. The
measured mean density of electrons $n_{\text e}$ was in the limits
(2.5--7.2)~$ \times 10^{10}$~cm$^{-3},$ the mean radius of grains
$r_{\text p} = 4 \times 10^{-5}$~cm, and the lower experimental
bound of the mean charge $Z \approx 500$.

For the grains with the indicated radius and the density $n_{\text
p} = 2 \times 10^6$~cm$^{-3},$ the dimensionless radius of a cell
$\varsigma \approx 62$. At the temperature $T = 2000$~K, the mean
charge and the mean density of thermoemission electrons are,
respectively, $Z \approx 442$ and $n_{\text e} \approx 6.91 \times
10^9$~cm$^{-3}$ according to \eqref{eq:charge_einb+cell} and
\eqref{eq:neutral_eq_3}; the charge distribution variance is
$Z_{\text D} \approx 49$ according to \eqref{eq:dispersion}. At the
temperature $T = 2200$~K, we have determined $Z \approx 560$,
$n_{\text e} \approx 8.76 \times 10^9$~cm$^{-3},$ and $Z_{\text D}
\approx 55$.

The calculated mean charge is close to the experimental result
($Z\approx 500$). However, the calculated densities of electrons are
less than experimental ones by one order. This fact can be
apparently explained by the neglect of the influence of an ionized
buffer gas in the calculations. Therefore, the comparison of the
theoretical and experimental results is not quite proper.

The study of the influence of the screening of a buffer gas will be
performed separately.

We note also that the essential excess of the mean charge $Z$ over
$Z_0$ is accompanied by the violation of the condition of
applicability of the linearized Poisson equation. The error of the
calculated mean charge $Z$ increases with $Z$, because, in this
case, the contribution of the region, where the nonlinear effects
are of importance, increases as well. This will be studied in
further work.

\vskip3mm The author thanks Prof. M.P. Malomuzh and Dr.~V.I.~Zasenko
for the fruitful discussion of the results. The present work was
partially supported by the Ministry of Education and Science, Youth
and Sport of Ukraine (grant No. 0112U001739909).

\vspace*{2mm}
\subsubsection*{APPENDIX\\ Electric Field of a Weakly Charged Spherical\\ Grain
in a Cell}\label{sec:appendix}

{\footnotesize Small charges $Z$ of a dust grain correspond to the
condition $r_{\text D} \gg r_{\text c} - r_{\text p}$. In this case,
the bulk density $\rho$ of thermoemission electrons in a cell has a
constant value $\rho = -en_{\text e}$, where $n_{\text e}$ is the
mean density of electrons in a cell, which is determined by the
obvious realation
\begin{equation}\label{eq:ap_a_el_den}
n_{\text e} = \frac{3}{4 \pi} \frac{Z}{r_{\text c}^3 - r_{\text
p}^3}.\tag{A1}
\end{equation}

The potential $\phi$ satisfies the Poisson equation
\begin{equation}\label{eq:ap_a_pois}
\Delta_r \phi = 4 \pi e n_{\text e}.\tag{A2}
\end{equation}
In the dimensionless variables \eqref{eq:dimless_varz}, Eq. (A2)
supplemented by the boundary conditions \eqref{eq:bcond} has the
solution
\begin{equation}\label{eq:ap_a_potential}
\psi(\tilde r) = 1 + \frac{Z/Z_0}{\varsigma^3 - 1} \left[
\frac{\varsigma^3}{\tilde r} - \frac32 \varsigma^2 + \frac12 \tilde
r^2 \right].\tag{A3}
\end{equation}
To within the designations, this formula  coincides with that obtained
in \cite{gibson} in the case of the weak screening.

The strength and the energy of the electrostatic field of a weakly charged
grain in a cell are given, respectively, by the relations
\begin{equation}\label{eq:ap_a_el_field}
\tilde E(\tilde r) = \frac{Z/Z_0}{\varsigma^3 - 1} \left(
\frac{\varsigma^3}{\tilde r^2} - \tilde r \right),\tag{A4}
\end{equation}
\begin{equation}\label{eq:ap_a_energy}
\tilde W_{\text{el}} = \frac12 \frac{Z^2/Z_0}{(\varsigma^3 - 1)^2}
\left[ \varsigma^6 - \frac95 \varsigma^5 + \varsigma^3 - \frac15
\right].\tag{A5}
\end{equation}
We note that, in the limiting case of small charges
(without the screening), we have strictly $\tilde E( \varsigma) = 0$.

}

\end{document}